\documentclass{aa}  

\usepackage{txfonts}

\usepackage{graphics,graphicx}
\usepackage{xcolor}

\usepackage{rotating,amssymb,amsmath}
\usepackage{soul}
\usepackage{booktabs}

\usepackage[colorlinks]{hyperref}
\hypersetup{colorlinks= true, citecolor= blue  }

\usepackage[english]{babel}
\usepackage{natbib}
\usepackage{parskip}
\usepackage[english]{babel}
\usepackage{morefloats}
\usepackage{color}
\usepackage{sidecap}
\usepackage{epsfig,color}
\usepackage{wrapfig}
\usepackage{lscape}
\usepackage{sidecap}
\usepackage{comment}
\usepackage{graphicx}

\newcommand{\CO}{\rm CO}

\newcommand{\Td}{\rm ${\rm T_{dust}}$}
\newcommand{\Tk}{${\rm T_{kin}}$}
\newcommand{\LII}{\rm ${\rm L^\prime}$}
\newcommand{\LIIunits}{\rm ${\rm K~ km~s^{-1}~pc^2}$}
\newcommand{\HII}{$\rm H_2 $ }

\newcommand{\My}{${\rm M_\odot yr^{-1}} $}

\newcommand\coone{$^{12}$CO(1-0)}
\newcommand\cothree{$^{12}$CO(3-2)} 
\newcommand\cofour{$^{12}$CO(4-3)}
\newcommand\cofive{$^{12}$CO(5-4)}
\newcommand\coseven{$^{12}$CO(7-6)}
 
\newcommand\cione{[{C}{I}](1-0)} 
\newcommand\citwo{[{C}{I}](2-1)}

\usepackage{xcolor}
\definecolor{ForestGreen}{RGB}{34,139,34}
\def\zy#1 {\textcolor{ForestGreen}{zy: #1} }

\begin{document}

   \title{Molecular gas excitation in the circumgalactic medium of MACS1931-26 }


   \author{L. Ghodsi
          \inst{1,2}\fnmsep\thanks{Contact email: layaghodsi@phas.ubc.ca}
          , 
          J. Zhou
          \inst{3,4}
          ,
          P. Andreani
          \inst{1}
          ,
          C. De Breuck
          \inst{1}
          , 
          A. W. S. Man
          \inst{2}
          , 
          Y. Miyamoto
          \inst{5}
          ,
          T.~G. Bisbas
          \inst{6}
          ,
          A. Lundgren
          \inst{7}
          ,
          Z.-Y.~Zhang
          \inst{3, 4}
          }

   \institute{ESO, Karl Schwarzschild strasse 2, 85748 Garching, Germany
         \and
             Department of Physics \& Astronomy, University of British Columbia, 6224 Agricultural Road, Vancouver BC, V6T 1Z1, Canada
         \and
             School of Astronomy and Space Science, Nanjing University, Nanjing, China
        \and
            Key Laboratory of Modern Astronomy and Astrophysics, Nanjing University, Ministry of Education, Nanjing, China
         \and
             Fukui University of Technology, 3-6-1, Gakuen, Fukui City, Fukui Pref. 910-8505, Japan
         \and
             Research Center for Astronomical Computing, Zhejiang Laboratory, Hangzhou 311100, China 
        \and
            Aix Marseille Univ, CNRS, CNES, LAM, F-13388 Marseille, France
             }

   \date{Received ......; accepted .....}

\titlerunning{CGM in MACS1931-26}
\authorrunning{Ghodsi et al.}


  \abstract
{The evolution of galaxies is largely affected by exchanging material with their close environment, the circumgalactic medium (CGM). In this work, we investigate the CGM and the interstellar medium (ISM) of the bright central galaxy (BCG) of the galaxy cluster, MACS1931-26 at $z\sim$0.35.
We detected \citwo, \coone, and \coseven\ emission lines with the APEX 12-m and NRO 45-m telescopes. We complemented these single-dish observations with \coone, \cothree, and \cofour\ ALMA interferometric data and inferred the cold molecular hydrogen physical properties. Using a modified large velocity gradient (LVG) model, we modelled the \CO\ and CI emission of the CGM and BCG to extract the gas thermodynamical properties, including the kinetic temperature, the density, and the virialisation factor. Our study shows that the gas in the BCG is highly excited, comparable to the gas in local ultra luminous infrared galaxies (ULIRGs), while the CGM is likely less excited, colder, less dense, and less bound compared to the ISM of the BCG. The molecular hydrogen mass of the whole system derived using \citwo\ is larger than the mass derived from \coone\ in literature, showing that part of the gas in this system is CO-poor. Additional spatially resolved CI observations in both transitions, \cione\ and \citwo, and the completion of the CO SLED with higher CO transitions are crucial to trace the different phases of the gas in such systems and constrain their properties.}
\keywords{galaxies: cluster - galaxies:evolution - submillimeter: gas}

\maketitle
%

\section{Introduction}

In the past decades, advances in observing techniques and galaxy formation simulations drew attention to the role of the circumgalactic medium (CGM) as a key ingredient to understand how material is exchanged between galaxies and their cosmological environment \citep{Werk14, Tumlinson17, Nelson20}. The CGM is defined as the multiphase gaseous halo extending from the outermost part of the galactic disc to the virial radius of the dark matter halo \citep{Tumlinson17}. For reference, the virial radius of a Milky-Way galaxy type is estimated to be $\sim\!200$ kpc \citep{Shull14}. Due to the diffuse nature of its gas, the exact extent of CGM is not straightforward to measure. However, observations of Neutral Hydrogen (HI) in $z<1$ massive galaxies hint a CGM extent of $\sim\!350$ kpc \citep{Wilde21}.
Inflows from the CGM to galaxies provide metal-poor gas as a fuel for galactic growth and sustaining star formation \citep{Voit21, Donahue22, Faucher23}. Such inflows are predicted by galaxy formation models \citep{Lochhaas20, Esmerian21}, however, they are difficult to be detected due to their extended spatial structure and faint emission. On the other hand, outflows powered by active galactic nuclei (AGNs) and star formation deposit metals back into the CGM \citep{Cheung16, Prochaska17, Christensen18}. These outflows can escape the potential well of their host galaxy or ``rain back'' onto the galaxies depending on their energy \citep{Voit21}. 
They can be detected more easily closer to their launch regions compared to when they arrive at the CGM because they are more compact at their launch regions. 
The physical conditions of the CGM could provide clues about potential inflows and outflows between a galaxy and its CGM \citep{Voit17}.

The CGM could be observed using both absorption and emission line spectroscopy. The warm and hot phase of CGM (T $>10^4$K) could be detected in absorption against bright background quasars mostly in optical and ultraviolet using neutral hydrogen and ionised lines of heavier elements \citep{Tumlinson11,Tumlinson17,Donahue22,Koplitz2023}. Ultraviolet surveys such as COS-Halos \citep{Tumlinson13, Werk16}, the Keck baryonic structure survey \citep[KBSS;][]{Rudie12, Rudie19}, and the cosmic ultraviolet
baryon survey \citep[CUBS;][]{Chen20, Zahedy21, Qu22} provide a rich sample of absorption line studies of the CGM of galaxies up to $z\sim 2$. 
Although this method is sensitive to a wide  range of densities, it only provides information about one line of sight towards the CGM and the background quasar.

The Ly-$\alpha$ transition of neutral hydrogen is the strongest emission from CGM \citep{Erb23}. This extended emission has been observed using the multi-unit spectroscopic
explorer (MUSE) at the ESO-very large
telescope (VLT) recently \citep{Wisotzki16, Wisotzki18, Chen21}. Despite the weak emission of $\rm H_2$ due to the lack of a permanent dipole moment, observing CGM and intergalactic medium (IGM) molecular gas is possible using the mid-infrared rotational $\rm H_2$ lines which trace the warm molecular gas (T $\sim100$K) partially \citep[$15\%-30\%$;][]{Togi16, Appleton17,Appleton23}.

The cold phase of $\rm H_2$ \citep[T $\sim10$K-$20$K;][]{Faucher23} is usually traced using other species, including CO and CI. CO is the second most abundant molecule after $\rm H_2$ and is a reliable proxy for molecular hydrogen in the local universe \citep{Bolatto13}. However, CO might fail in tracing all the $\rm H_2$ content in some cases due to the dependence on redshift, radiation field, metallicity, gas excitation, and the cosmic rays (CR) ionisation rate \citep{Bisbas+17,Valentino18, Madden20}. Specifically, CRs might be one of the main heating sources of gas in and around star-forming galaxies since CRs can penetrate deeply into molecular clouds and regulate the thermal balance of the gas \citep{Papadopoulos10, Grenier15, Gaches19, Bisbas+21}. CRs can destroy CO in molecular clouds and leave large swathes of CO-poor/([CI]-[CII])-rich gas \citep{Bisbas+15,Bisbas+17,Papadopoulos+18}. 
In these cases, a combination of CO and CI is a good tracer of the cold molecular gas content of the CGM.

Interferometers provide high-resolution observations of CGM that are crucial to study the structure and kinematics of this gas. However, interferometers cannot detect the emission of scales larger than their largest angular scale which is determined by the shortest baseline of the observing configuration \citep{Plunkett23}. Hence, due to the diffuse and extended ($\sim 200$kpc) nature of CGM, it is not straightforward for interferometers to detect the whole extent of CGM. Single-dish observation, on the other hand, are able to detect the emission of scales as large as their field of view \citep{ Lee24}. Combining the information obtained from both single-dish and interferometric observations is thus necessary to study the CGM.

On the observational front, recent studies have been able to detect CGM, IGM, and intracluster medium (ICM) using the CO and CI emission lines originating from cold gas. These streams seem to feed the star formation activity of galaxies as a part of the baryon cycle in the CGM. For instance, multiphase emission from the tidal streams in the IGM of a galaxy group, Stephan's Quintet, is detected at z=0.02 \citep{Appleton23}. Extended molecular structures are found in halos (r$\sim$10-20~kpc) of z$\sim$2.2 QSO \citep{Scholtz+23} and out to r$\sim$200~kpc \citep{Cicone+21}.
\citet{Emonts+23} find a \cione~stream around the massive $z$=3.8 radio galaxy 4C 41.17, likely linked to its strong star formation. CO and \cione~molecular winds are imaged in local luminous infrared galaxies \citep[LIRGs; ][]{Cicone+18}. These winds seem to be linked to  AGN  and star formation at a rate of SFR$\sim$~20-140 \My \citep {Cicone+14}. Stronger molecular outflows originating from galaxies in clusters with SFR$\sim$~150-250 \My can eject gas and dust further away from the CGM into the intracluster medium \citep{Fogarty19,Emonts+18}.

Extracting information about the physical conditions of the gas detected in the spectral lines of different species, requires both radiative transfer modelling and assumptions about the excitation of the gas \citep{VanderTak07, Roueff24}. The fastest and simplest framework for this problem is the local thermodynamic equilibrium (LTE) assumption which holds in high densities and states that the excitation temperature equals the kinetic temperature. However, this assumption might introduce biases to the line excitation modelling \citep{Roueff21}. Some non-LTE methods, including the large velocity gradient assumption \citep[LVG;][]{Sobolov60}, consider a local excitation, meaning that the species excitation is a result of their interactions with local radiation field. For instance, RADEX \citep{VanderTak07} is a non-LTE code, solving the radiative transfer and molecular excitation under the LVG assumption which is widely used in molecular excitation studies. More sophisticated radiative transfer methods can treat the problem in a non-local way \citep{Asensio18}. However, these models are computationally expensive and more complicated to implement \citep{Roueff24}.

In this work, we trace the molecular gas content of the BCG of a galaxy cluster, MACS1931-26 (hereafter MACS1931), and its CGM using a combination of CO and CI emission lines. Our goal is to model the thermodynamical state of the gas in CGM and BCG using a radiative transfer code and study the potential differences of gas in these two components of the galaxy.
The MACS1931 galaxy cluster is an excellent testbed for investigating the physical conditions of molecular gas in the ISM inside the BCG and its CGM due to its strong star formation activity and X-ray-loud AGN \citep[$\rm SFR \sim 250~ M_{\odot}yr^{-1}$;][]{Ehlert11, Santos+16, Fogarty17, Fogarty19}. MACS1931 is a cool-core galaxy cluster having one of the largest known \HII reservoirs
($\sim 2\times10^{10}\,M_{\odot}$) in a cluster core revealed by the Atacama Large Millimeter/submillimeter Array (ALMA) \cothree\ and \coone\ imaging \citep{Fogarty19, Rose19}. An appreciable fraction of this mass is found over a tail-like structure extending up to 30\,kpc from the BCG core \citep{Fogarty19}. This tail-like structure (hereafter tail) is a part of the multiphase CGM of the central galaxy and contains a  large reservoir of  $\rm T_{dust}\sim$10\,K dust.

Similar to the Phoenix cluster BCG \citep{Russell+17}, the BCG of MACS1931 and its tail-like structure have a low gas-to-dust ratio of $\sim10-25$ \citep{Fogarty19} lower than the Galactic value of $100-150$ \citep{Draine84, Remy14}. The unusually low ratio suggests that a potentially large \HII gas reservoir might have been missed by the existing CO observations, perhaps due to the CR-irradiation rendering some of the \HII gas CO-poor, and/or because ALMA long baseline interferometric observations might have filtered out extended \coone\ line emission \citep{Lee24}. To investigate this matter further, we obtained additional single-dish \coone\ observations to be able to recover the total flux, which might be filtered out in interferometric observations, along with high-frequency CO and CI observations 
in order to probe the gas physical conditions. 
While the detected CO emission with ALMA represents only part of the whole CGM\footnote{We call this tail-like structure a CGM component of the MACS1931 BCG since it is close to the galaxy, it is detected with cold molecular gas tracers, and there is no spectroscopically confirmed neighbour galaxy in the direction of the tail.} of MACS1931 BCG, observations with existing single-dish telescopes can help determine how much of the emission is over-resolved by ALMA. The detection of CO and CI emission at 30 kpc from MACS1931 offers a unique opportunity to study a cloud that is likely part of the CGM surrounding this BCG.

In this article, we report new \coone, \coseven~, and \citwo~single-dish observations towards the MACS1931 galaxy cluster at $z=0.3525$. We complement these data with the ALMA/ACA observations presented in \citet{Fogarty19} and use the combined dataset to provide an analysis of the average physical conditions of the BCG and its CGM in this galaxy cluster. 
Throughout, we adopt a $\Lambda$-CDM Cosmology as defined by \citet{Planck2016} where $\rm H_0 = 67.8~km~s^{-1}~Mpc^{-1}$ and
$\rm \Omega_m = 0.308$.


\section{The observations}\label{obs}

In this section, we report the methods and details of the observations of the \citwo, \coone, and \coseven\ lines with the APEX 12-m and the NRO 45-m antenna telescopes as well as the archival ALMA data towards MACS1931, used to complement our single-dish data.

\begin{figure*}
\hspace{0cm}
\includegraphics[width=18cm, height=14cm]{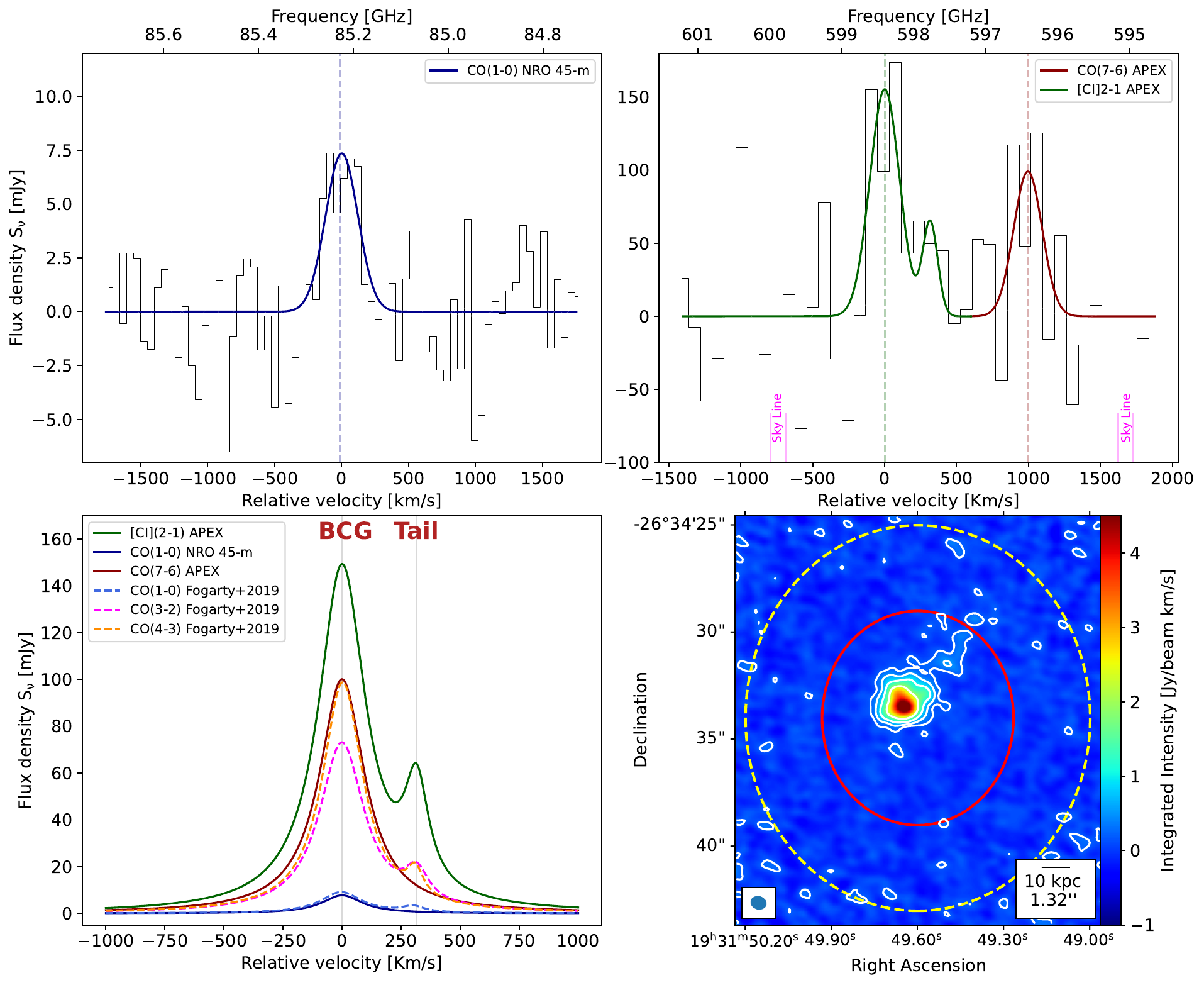}
\caption{Submillimeter observations towards MACS1931. \textit{Upper left:} Gaussian fit to the \coone~line in the NRO 45-m
observations of MACS1931. The black histogram shows the spectrum. The blue
solid curve shows the Gaussian fit to the line. \textit{Upper right:} Gaussian fit to the \citwo~and \coseven~lines in the APEX observations of MACS1931. The
black histogram shows the spectrum. The green and red solid curves show the
Gaussian fits to the  \citwo~and \coseven~lines, respectively. The magenta solid lines show the regions of the spectrum excluded because of the atmospheric lines. \textit{Lower left:} Best-fit line profiles of \coone, \cothree, and \cofour~from \citet{Fogarty19} (dashed curves) and
Lorentzian fits on \coone, \coseven, and \citwo~from this work (solid
curves) displayed on the same plot for comparison (see \S\ref{results} and
Table~\ref{tab:fluxes}). The CGM tail peak of the \cofour~line from \citet{Fogarty19} is the best fit to a partially truncated spectrum due to a spectral window gap in the ALMA Band 7 observations. \textit{Lower right:} The ALMA \cothree~ intensity map of the MACS1931 BCG and the tail. The moment zero map is calculated over a range of $[-500,600]\ \rm km\,s^{-1}$. ALMA beam is shown by an ellipse on the bottom left side of the image. The APEX and NRO 45-m HPBWs are shown by solid red and dashed yellow circles. Contours are calculated on the moment map with levels $[0.2, 0.5, 1, 2]\sigma$ and $\sigma=0.297 \ \rm Jy\,beam^{-1}.km\,s^{-1}$.
The CGM tail component of the spectra is detectable in the \citwo~line as well as all lines from \citet{Fogarty19}.}
\label{fig:all_obs}
\end{figure*}

\begin{table*}
 \centering
    \caption{Line flux and luminosity towards MACS1931.} 
    \begin{tabular}{l c c c c c c c c c }
 \hline
 \bf{Line} & \bf{Component} & \bf{Velocity} & \bf{Method}& \bf{FWHM}  & \bf{Flux}& \bf{Flux$\rm{_{beam \ corr}}$} & \bf{Luminosity L$^\prime$} & \bf{L$^\prime$$\rm{_{beam \ corr}}$} & \bf{S/N} \\
 && [$\mathrm{km ~s^{-1}}$] && [$\mathrm{km~ s^{-1}}$] & \multicolumn{2}{ c }{[ \ \ $\mathrm{Jy~ \ \ \ km ~s^{-1}}$ \ \ ]} & \multicolumn{2}{ c }{$\mathrm{10^9 \ \ \ [ \ \ K \ \ kms^{-1} \ \ pc^{-2}} \ \ ]$} & \\
 \hline
 \bf{CO(1-0)}&\bf{BCG} &  fixed $0$  &Gauss& $  285 \pm 54$ & $  2.2\pm 0.6 $&- &$9.8 \pm 2.4 $  & - & 4.2 \\
  &&&Lorentz&$207 \pm 32$ &$  2.5 \pm 0.6 $& &$11.0 \pm 2.5$&  & \\
 &\bf{Tail} &  -  & Gauss& fixed 123 & $<$0.8  & $<0.9$& $<3.7$  & $<4.3$&-\\
\hline
\bf{CO(7-6)}&\bf{BCG} & fixed  &Gauss& $233 \pm 43$ & $24.4 \pm 10.3$   &-  & $2.2 \pm 0.9 $   &- & 2.4\\
 &&$994.7 $&Lorentz& $298 \pm 83$ &$36.7 \pm 13.5$& &  $3.3 \pm 1.2$&  & \\
 &\bf{Tail} & - &Gauss& fixed 123 & $<$27.9 & $<47.1$ &  $<2.5$  & $<4.2$ & -\\
\hline
\bf{[CI](2-1)}&\bf{BCG} &  fixed $0$ &Gauss& $243 \pm 103$ & $41.6  \pm 13.2$ & - & $3.7 \pm 1.2$ & - &   3.7\\
 &&&Lorentz&$245 \pm 174$&$56.4  \pm 14.7 $ & &$5.0\pm 1.3$& &  \\
 &\bf{Tail} & fixed   &Gauss& fixed 123  & $ 8.4 \pm 6.0$  &$14.2\pm10.1$ & $0.7 \pm 0.5$  & $ 1.3\pm 0.9$&  1.2\\
 &&$315.8 $&Lorentz&&$ 8.6 \pm 7.3$ &$14.5 \pm 12.3$ &$0.8 \pm 0.6$& $ 1.3\pm 1.0$ & \\
\hline
    \end{tabular}\label{tab:fluxes}
    \vskip 0.2cm
    \noindent
\tablefoot{In this table, "$<$" symbols refer to estimated $3\sigma$ upper limits.\\ 
For each spectral line of each component of the galaxy (Tail and BCG), the velocity offset and the full width at half maximum (FWHM) of the fitted function are reported in the third and fifth columns of the table, respectively. The FWHM of all lines for the tail are assumed to be the same as the FWHM of the \cothree~ line detected by \citet{Fogarty19} for the tail of MACS1931. \\
Line fluxes are computed by two methods of fitting the line profile with a Gaussian function and a Lorentzian function. The measured flux and luminosity using each method are reported in the sixth and eighth columns, respectively.\\ 
"Beam corrected" fluxes and
luminosities are corrected for the differential beam effect and reported in the seventh and ninth columns, respectively.\\
The signal-to-noise ratio (S/N) of the Gaussian fitting function for each spectral line is reported in the tenth column.}
    \end{table*}

\subsection{APEX observations}

We used the Atacama Pathfinder EXperiment (APEX) telescope \citep{Gusten06} equipped with the SEPIA660 receiver \citep{Belitsky18} tuned at the frequencies of the \coseven~and \citwo~lines (central frequency of 609.5 GHz). The observations of MACS1931-26 for this project were obtained from 2021 June 25 to 2023 April 27 (project ID: E-0106.A-0997A-2020, PI: P.Andreani). The data taken in April 2023 did not meet good quality standards and were omitted from the analysis.

The APEX/SEPIA660 receiver is a single-pixel heterodyne receiver working at the frequency window of 597-725 GHz. The total integration time was 20 hours, centring towards the BCG on the direction of $\mathrm{RA = 19^h31^m49^s.60}$ $\mathrm{Dec = -26^\circ34'34''}$. The precipitable water vapour (PWV) was 0.4-0.9 mm on June 25, 0.2-0.8 mm on June 28, and 0.1-0.4 mm on June 29. The intended spectral lines, \citwo~with a rest frequency of 809.34 GHz and \coseven~with a rest frequency of 806.65 GHz, are redshifted to the frequencies of 598.40 GHz and 596.41 GHz at the redshift of z=0.3525 \citep{Fogarty19}, respectively. The wobbler mode used in these observations had an amplitude of $50''$ and a frequency of 1.5 Hz. The half-power beam width (HPBW) of APEX for this observation is $9.89^{\prime\prime} \pm 0.11^{\prime\prime}$.

We reduced the data using the GILDAS/CLASS\footnote{\url{https://www.iram.fr/IRAMFR/GILDAS/}} \citep{Pety05}
software package. Two parts of the APEX spectrum near 595 GHz ($\rm O_3$) and 600 GHz ($\rm O_3 + HDO$) are affected by atmospheric absorption \citep{Pardo22}. For each scan, we excluded $\sim0.3$ GHz of the spectrum at the frequency of each atmospheric line. A first-order baseline was then fitted to each scan and subtracted from the spectrum. Finally, we averaged the spectra of all scans and fitted Gaussian functions for the \coseven~and \citwo~emission lines. The RMS noise of this spectrum is 0.66 mK per channel with channel width of 82 $\rm kms^{-1}$, measured using CLASS on continuum-free line-free channels.
No planet was available during the observations. Instead, calibration observations were performed on the source IRAS15194-5115. These reference spectra were compared to the calibration observations on the same source in performed in mid 2022. The flux ratio of this calibrator source was applied to cross scans of Mars obtained in 2022 June with SEPIA660, yielding the antenna gain of \footnote{ Conversion factor $\rm \Gamma [Jy/K]$=$\rm S_{\nu}/T^{*}_A$=$\rm 8
k_B/(\eta _A \pi D^2)$  where $S_{\nu}$ is the flux density; $T^{*}_A$ is the brightness temperature; $\eta _A $ is the aperture efficiency; and $D$ is the antenna diameter.} $\Gamma$=(72 $\pm$ 8)Jy/K for 2021 which we used to convert the observed brightness temperature to flux density.

After data reduction within GILDAS/CLASS, we performed a Monte Carlo simulation using the Astropy and Specutils Python packages \citep{astropy22, specut23} to
estimate the uncertainty of these measurements. In each realisation, we added an artificial Gaussian error 
to the data with a standard deviation equal to the RMS of the data (47 mJy per channel).
Then, we fitted Gaussian functions to both lines and measured all of the line properties in each realisation. The reported values in Table~\ref{tab:fluxes} are the median and standard deviation of these
parameters.


\subsection{NRO 45\,-m dish observations}

The \coone\ single-dish observations towards MACS1931 were taken with the 45-m telescope of the Nobeyama radio observatory (NRO) in Japan, which is equipped with the FOur-beam REceiver front-end \citep[FOREST;][]{Minamidani+2016}.  FOREST is a four-beam (2$\times$2)  dual-polarization sideband-separating SIS mixer receiver for the single sideband (SSB) operation.
\coone\ has a rest frequency of 115.27 GHz that is redshifted to 85.22 GHz at $z=0.3525$.
The frequency coverage and the resolution for each array are set to 1 GHz and
244.14 kHz, respectively, giving a velocity coverage and resolution of $\sim$3500 ${\rm km s^{-1}}$ and 0.86 ${\rm km s^{-1}}$ at 85.22 GHz. At this frequency, the HPBW of the 45-m telescope with FOREST is $\sim$18$^{\prime\prime}$. 
We adopted the ON-ON observation mode, which uses two of the four beams of the FOREST receiver at the diagonal (beams 1 and 3).
Since the two beams are separated by 70'' in the sky, one of the beams observes the object while another the sky when the apparent size of the object is smaller than the beams. The dominant observation overhead is therefore the slewing time for beam switching.
We adopted a switching period of 10s cycle to optimize the observation overhead and the bandpass stability \citep{Nakajima}.

The data were taken on 2022 March 1, 2, and 12.
The observing time for each day was roughly 5 hours, while the on-source time was 3.8, 6.4, and 5.8 hours (total hours for beam 1 and beam 2, and H- and V-polarisation). The system temperature was 255\,K, 247\,K and 211\,K for each day, respectively.
In order to check the absolute pointing accuracy, we observed a point source (GL2370), which is the closest and strongest SiO maser source to the target using a 43 GHz band receiver every hour.
The reported antenna temperature was obtained by the chopper wheel method, yielding the $\rm T_A^\star$ scale, which contains corrections for atmospheric and any antenna Ohmic losses as well for beam pattern rearward spillover and losses \citep{Kutner&Ulich}. The main beam brightness temperature, $\rm T_{mb}$ was converted from  $\rm T_A^\star$, for each intermediate frequency by observing a standard source, the carbon star IRC+10216. The scaling factors take into account the main-beam efficiency ($\eta _{\rm mb}$) of the 45-m antenna, but also compensate for the decrease in line intensity due to the imperfect SSB image rejection.

The reduction of the NRO 45-m data towards MACS1931 was carried out using the NEWSTAR software \citep{Ikeda01} which is a data reduction package designed for this telescope. After flagging out the bad spectra, we integrated the spectra for each schedule. A first-order baseline was then subtracted from the spectra and the intensities were corrected with a point-source Jy/K conversion factor of $\Gamma $=($6.54 \pm 0.33$)\,Jy/K. Finally, we integrated the spectra over all schedules and smoothed the resulted spectrum to a velocity channel of bin width 50 $\rm{km\,s^{-1}}$.
We derived an aperture efficiency of $\eta_A$$\sim$$0.28\pm0.06$, appropriate for the observed frequency and elevation 
from the information provided by the observatory\footnote{\url{https://www.nro.nao.ac.jp/~nro45mrt/html/prop/eff/eff2022.html}}. 
The source declination allowed us to conduct observations at elevations $\sim 30^{\circ}$ above the horizon. The source was detected in each observing run, however, it was heavily affected by the atmospheric absorption with an RMS noise of 2.6 mJy per $50 \ \rm{km\,s^{-1}}$ channel. The line properties and uncertainties were calculated using the same method as the APEX data (Table~\ref{tab:fluxes}).


\subsection{Archival ALMA data}

We used archival ALMA data towards MACS1931 published by \citet{Fogarty19} to complement our single-dish data. The observations were obtained in ALMA Band 3, cycle 4 (Project ID: 2016.1.00784.S, PI: M.Postman) and Band 6 and 7, cycle 5 (Project ID:
2017.1.01205.S, PI: M.Postman). The Band 3 data is a combination of two ALMA 12-m array configurations with a total integration time of 13366 s, minimum synthesised beam of 0.54$^{\prime\prime}$ and maximum recoverable scale of $18.8^{\prime\prime}$. The Band 6 data towards MACS1931 is a combination of an ALMA 12-m array configuration with 726 s integration time and an Atacama compact array (ACA) configuration with an integration time of 1724 s. The  minimum synthesised beam and maximum recoverable scale are 0.74$^{\prime\prime}$ and 20.4$^{\prime\prime}$, respectively. Finally, the Band 7 data is a combination of the 12-m array (integration time of 2722 s) and ACA (integration time of 6804 s), yielding minimum synthesised beam and maximum recoverable scale are 0.81$^{\prime\prime}$ and 14.6$^{\prime\prime}$, respectively \citep{Fogarty19}. We used the CO line measurements of \citet{Fogarty19} to complement our data and model the gas excitation.


\section{Results}\label{results}

In this section, we present and analyse the results of the APEX and NRO 45-m observations of MACS1931 combined with the archival ALMA data. These single-dish observations do not have the required spatial resolution to distinguish between the BCG and the tail of MACS1931 and we use the ALMA observations of \citet{Fogarty19} as a prior to understand the molecular gas geometry and kinematics. However, we are able to detect the tail emission spectrally using the \citwo~ line with APEX.

    \begin{table}\label{tab:lines}
 \centering
    \caption{Line luminosity ratios towards MACS1931-26.}   
    \begin{tabular}{l c c }
\hline
\bf{Line luminosity ratio} & \bf{BCG} & \bf{Tail} \\
\hline
\bf{CO(3-2) / CO(1-0)} &$0.97 \pm 0.09$ & $0.67\pm 0.13$ \\
\hline
\bf{CO(4-3) / CO(1-0)} &$0.67 \pm 0.07$ & $0.22 \pm 0.11 $ \\
\hline
 & $0.17 \pm 0.07$& \\
 \bf{CO(7-6) / CO(1-0)} & ($0.25 \pm 0.09$ & - \\
 &  for Lorentzian) & \\
\hline
 & $0.28 \pm 0.09$&$ 0.68\pm 0.48$ \\
\bf{[CI](2-1) / CO(1-0)} & ($0.38 \pm 0.09$ & ($0.68 \pm 0.53$ \\
 & for Lorentzian) & for Lorentzian) \\
\hline
    \end{tabular}
    \vskip 0.2cm
    \noindent
\label{tab:lines}
    \tablefoot{All line ratios are calculated using the ALMA interferometry \coone~\citep{Fogarty19} because we do not detect the tail \coone~peak in single-dish observations. Moreover, the \cothree~and \cofour~fluxes are obtained from \citet{Fogarty19}.}
    \end{table}

We present the detected \CO\ and CI lines in Figure~\ref{fig:all_obs}. The upper left panel shows \coone ~spectrum obtained with NRO 45-m/FOREST receiver. The upper right panel shows the \citwo\ and \coseven\ spectrum obtained with the APEX/SEPIA660 receiver.
Despite the atmospheric noise affecting the APEX high-frequency data and the NRO 45-m low elevation data of MACS1931, the lines are detected, albeit at low statistical significance.
To extract the line properties over an integrated region as large as the HPBW ($\sim 10^{\prime\prime}$ and $\sim 18^{\prime\prime}$ for APEX and NRO 45-m, respectively), we used a Lorentzian fitting function as well as a Gaussian function. We employed the Lorentzian fits because in the following we compare the fluxes extracted from these single-dish observations with those reported by \citet{Fogarty19} using a Lorentzian fitting function. 
As described in \citet{Fogarty19}, the spectrum of each line has a bimodal distribution including a larger peak corresponding to the BCG and a more redshifted smaller peak corresponding to the tail. Because of the low S/N in single-dish observations, this bimodal distribution is only detectable in the \citwo~spectrum (Figure \ref{fig:all_obs}). As a reference for the line properties, we used the second velocity component (C2) of $\mathrm{^{12}CO(3-2)}$ reported in \citet{Fogarty19} corresponding to the 30 kpc tail-like structure since this line has the highest S/N in their work. Assuming that the velocity offset between the two peaks and the FWHM of the small peak are the same as the $\mathrm{^{12}CO(3-2)}$ line from \citet{Fogarty19}, we extracted the line properties, including line widths, fluxes, and luminosities, reported in Table~\ref{tab:fluxes}.

The best-fit parameters of these models are listed in Table~\ref{tab:fluxes} together with uncertainties resulting from the calibration, data reduction, and error estimation process described in the previous section.
The lower right panel of Figure \ref{fig:all_obs} shows a comparison between the beam sizes of ALMA, APEX, and NRO 45-m observations. Based on this figure, the tail is distributed from the BCG (the centre of the field of view) up to $\sim6^{\prime\prime}$ away from the centre. Hence, a fraction of the tail flux is not captured by the APEX and NRO 45-m telescopes due to the beam response.
A Gaussian beam response function was assumed to calculate the "beam corrected" fluxes and luminosities in Table~\ref{tab:fluxes}. 
The exact extent of the tail-like CGM component is not measurable and the emission is not spatially resolved in these single-dish observations. Therefore, in order to account for the Gaussian beam response, we assumed the tail to be a point source at a distance of $\sim 4^{\prime\prime}$ from BCG and the beam centre and increased the tail flux by a fraction for the beam response correction. Considering that the APEX HPBW is 9.89$^{\prime\prime}$, the tail flux drops by $\sim 40\%$ at the distance of 4$^{\prime\prime}$ in a Gaussian beam. So, we added this fraction to the tail flux of \citwo~ measured before. We also added this fraction to the upper limit flux of \coseven~measured from APEX data. Through a similar analysis for NRO 45-m dish with a HPBW of 18$^{\prime\prime}$, we added $\sim 15\%$ to our upper limit estimation of the tail flux in \coone. 

From the spectra of the \coseven\ and \coone\ lines, we determined the line flux of the BCG and extract upper limits on the tail flux values. We report the $3 \sigma$ upper limits for these lines (see Table~\ref{tab:fluxes}) by computing $\sigma = \mathrm{RMS \sqrt{N_{\rm chan}}  \delta v}$, where $\mathrm{RMS}$ is the noise per channel of the spectrum, $\mathrm{N_{chan}} = 2\mathrm{FWHM_{CO(3-2)}/\delta v}$ is the number of line channels, and $\mathrm{\delta v}$ is the channel width (82 $\rm kms^{-1}$ and 50 $\rm kms^{-1}$ for APEX and NRO 45-m, respectively).
The lower left panel of Figure~\ref{fig:all_obs} shows the Lorentzian line models for all the CI and CO lines detected for MACS1931 reported in this work and \citet{Fogarty19} for comparison.
The line luminosity ratios, in units of \LII\ [\LIIunits], inferred for MACS1931 for all the reported lines for the BCG and tail components are listed in Table~\ref{tab:lines}. Line luminosities are computed following \citep{Solomon97}:

\begin{equation}
    L^{\prime}_{line} = 3.25 \times 10^7 S_{line} \Delta V \nu^{-2}_{obs}D^2_L (1+z)^{-3}
    \label{Lprime}
\end{equation}

\noindent
where $S_{line}\Delta V$ is the integrated line flux; $\nu_{obs}$ is the redshifted frequency; $D_L$ is the luminosity distance; and $z$ is the redshift of the object.

Figure \ref{fig:all_obs} shows that the \citwo~line is brighter than the \coseven~line and the full width at half maximum (FWHM) is similar for both lines within the error bars. This is indicative of a highly excited warm gas rich in atomic Carbon.
The single-dish flux value of the \coone~line ($2.5 \pm 0.6 \ \rm{Jykms^{-1}}$ using a Lorentzian fit) is in agreement with the one detected with ALMA 12-m array reported in \citet{Fogarty19} ($3.02 \pm 0.42  \ \rm{Jykms^{-1}}$) within the error bars. It would imply that the ALMA interferometric observations do not filter out flux from an extended component of this system. Notice that the maximum recoverable angular scale in Band 3 for the observations reported by \citet{Fogarty19} is 18.8$^{\prime\prime}$ is almost equal to the half-power beam width of the NRO 45-m observations (see \S~\ref{obs}). Consequently, missing \coone\ flux in ALMA observations is not the main reason for the low gas-to-dust ratio of MACS1931. We investigate further whether this system contains some CO-poor gas seen using different tracers.


\subsection{Spectral line energy distribution and profiles}

\begin{figure*}
\hspace{0.7cm}
\includegraphics[width=16cm]{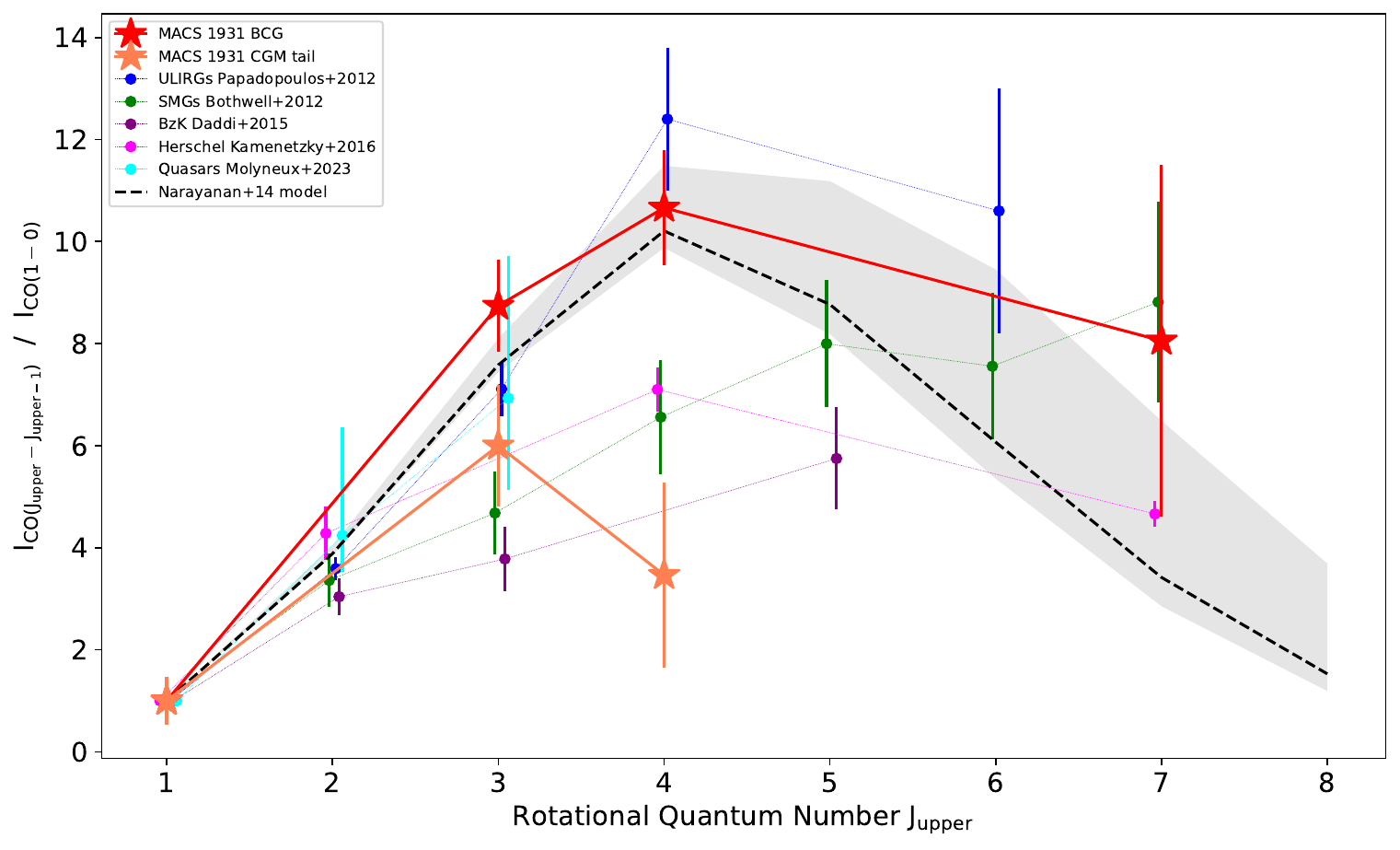}
\caption{CO SLED built with the available data for the BCG and the CGM tail of MACS 1931 (red and orange stars, respectively). The CO line fluxes are taken from \citet{Fogarty19}, except \coseven~of the BCG which is detected in this work using APEX. Line fluxes are normalised to the \coone~ flux taken from \citet{Fogarty19}. Additional average CO SLEDs from the literature are added for comparison: ULIRGs from \citet[][blue circles]{Papadopoulos+12}, SMGs from \citet[][green circles]{Bothwell13}, BzK galaxies from \citet[][purple circles]{Daddi15}, low-redshift Herschel galaxies from \citet[][magenta circles]{Kamenetzky16}, and quasars from the QFeedS sample published in \citet[][cyan circles]{Molyneux23}. The dashed black line shows the theoretical CO SLED for resolved objects with $\Sigma_{\rm SFR}=0.83 \ M_{\odot} \rm yr^{-1} \rm kpc^{-2}$ modelled by \citet{Narayanan14}. This value is the estimated $\Sigma_{\rm SFR}$ for the MACS1931 BCG \citep{Fogarty15}. The grey shaded area shows the predicted CO SLED range with this model for $0.6< \Sigma_{SFR} \ [M_{\odot} \mathrm{yr}^{-1} \mathrm{kpc}^{-2}] < 3$.
We report an upper limit on the \coseven~flux for the tail in Table~\ref{tab:lines}, however, it is not constraining the tail CO SLED and falls out of the range of this plot.}
 \label{COSLED}
\end{figure*}

 The new observations reported in this work help to further constrain the physical state of the gas content of this system. Figure~\ref{COSLED} shows the normalised CO flux ratios against the rotational quantum number, the so-called CO spectral line energy distribution (CO SLED), of the BCG and tail separately.  
 The line luminosities are normalised to the interferometric \coone\ line values obtained by from \citep{Fogarty19} for both BCG and tail.
In the CO SLED of the tail, we do not display the \coseven/\coone\ ratio because we only have an upper limit of the \coseven\ flux that is not stringent enough to constrain the CO SLED.
In Figure \ref{COSLED}, the CO SLED of MACS1931 is compared to those of local ULIRGs \citep{Papadopoulos+12}, submillimetre galaxies \citep{Bothwell13}, BzK galaxies \citep{Daddi15}, quasars \citep{Molyneux23} and local IR galaxies observed by Herschel \citep{Kamenetzky16}.
The line flux ratios of the MACS1931 BCG are particularly large similar to those found in local ULIRGs \citep{Papadopoulos+12,Montoya-Arroyave+23}. These large values are characteristics of high excitation, likely ascribed to starburst gas. On the contrary, the line flux ratios related to the tail component of MACS1931 do not show the same trend as the BCG, implying that the excitation of CO in the tail is different from that in the BCG and in ULIRGs. Moreover, the \cofour/\cothree\ ratio is relatively low which differs from that measured in Milky Way-like galaxies \citep[Figure \ref{COSLED};][]{Kamenetzky16}. It is noteworthy that the ALMA Band 7 spectrum of MACS1931 is truncated around the relative velocity of $300-500 \ \rm kms^{-1}$ for the \cofour\ line, and the reported line flux in \citet{Fogarty19} is based on the best fitted Lorentzian function on this partially obscured spectrum. Hence, the actual \cofour~flux might be different from the reported value. More observations of different \CO~lines, especially higher transitions like \cofive, would be beneficial to better constrain the CO SLED of the tail. We put models into practice to interpret these results.

\subsection{Line ratio modelling}\label{modeling}

\begin{figure}
     \centering
         \includegraphics[trim=0 0 0 0,clip, width=0.5\textwidth]{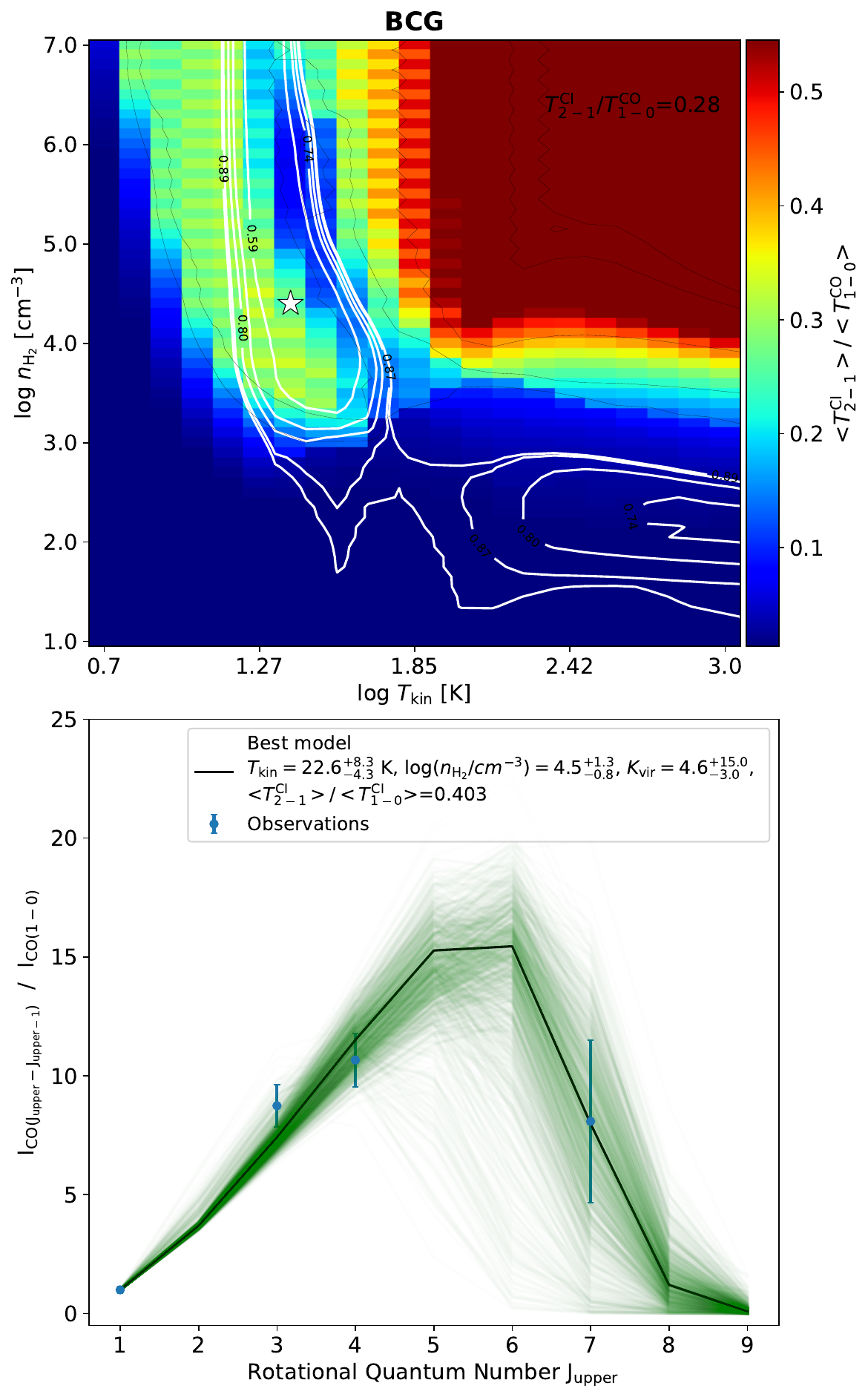}
        \caption{\textit{Top:} LVG modelling for the \textbf{BCG} of MACS1931. The \Tk, $\rm n_{H_2}$ grid is plotted with a projected $\rm K_{\rm vir}$ axis. At each point of the \Tk, $\rm n_{H_2}$ grid, the colour shows the predicted \citwo/\coone\ ratio of the model with the highest likelihood among models with different $\rm K_{\rm vir}$ values. The black contours show the levels of the modelled \citwo/\coone\ flux ratio which is shown by colours. The white contours show the likelihood levels. The white star indicates the best-fitted LVG model obtained from MCMC simulations. \textit{Bottom:} The CO SLED of the best-fitted LVG model for \textbf{BCG}. The black curve shows the modelled CO SLED. The black points with blue error bars show the observed line flux ratios. The green curves show the distribution of best models obtained from MCMC. 
        }
        \label{fig:BCG_model}
\end{figure}

\begin{figure}
     \centering
         \includegraphics[trim=0 0 0 0,clip, width=0.5\textwidth]{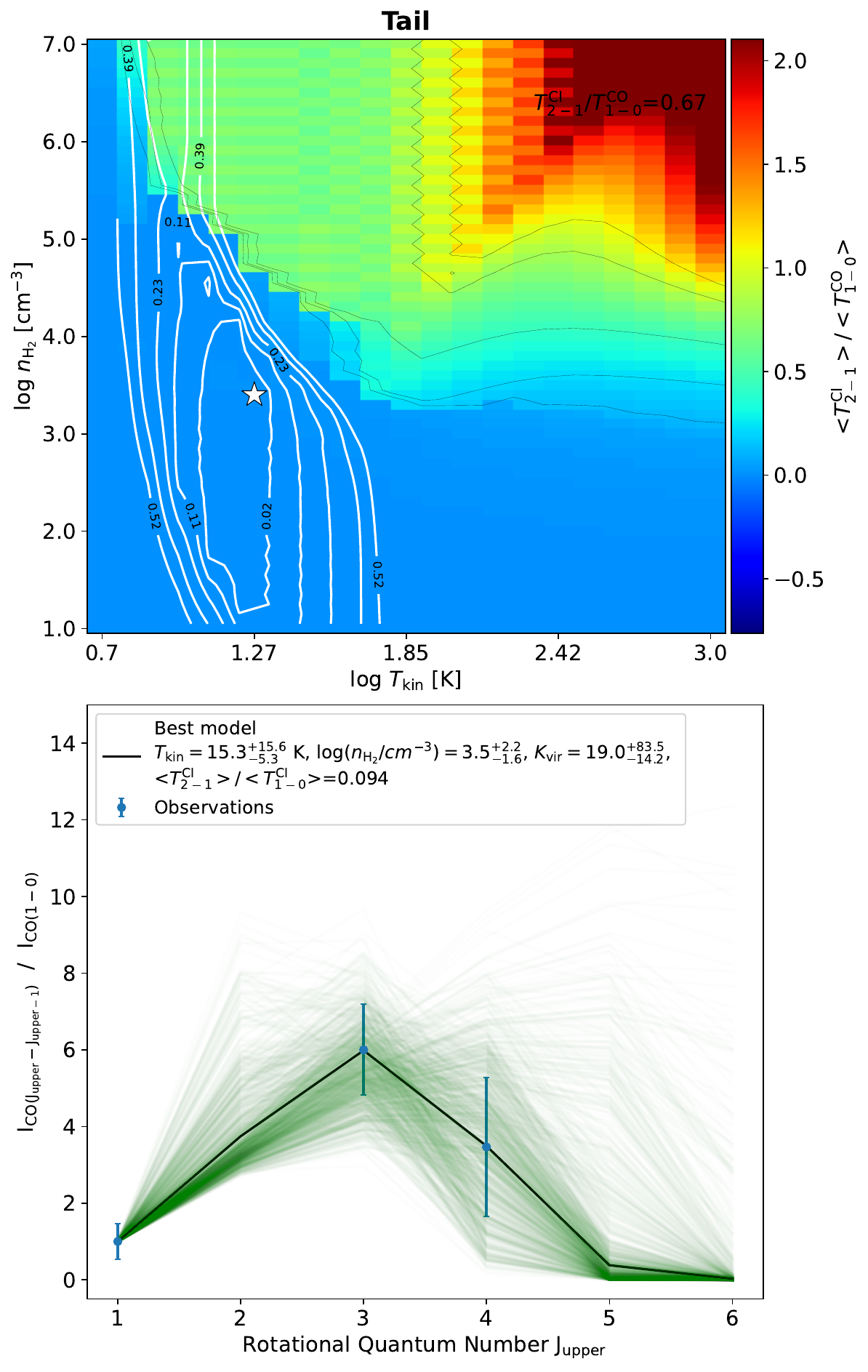}
        \caption{LVG modelling and CO SLED for the \textbf{CGM tail} of MACS1931, similar to Figure \ref{fig:BCG_model}.}
        \label{fig:CGM_model}
\end{figure}

The inference of the physical state of gas from the molecular line luminosities is not straightforward. We can constrain the gas thermodynamical properties using the observed line fluxes with the help of reasonable physical assumptions. To this aim, we used a non-local thermodynamic equilibrium (non-LTE) radiative transfer code, i.e.
MyRadex\footnote{\url{https://github.com/fjdu/myRadex}} \citep{Du2022} to map the [$n_{H_2}$, $T_{\rm kin}$, $K_{\rm vir}$] parameter space for the CO lines. This parameter space includes the molecular hydrogen number density ($n_{H_2}$), kinetic temperature ($T_{\rm kin}$), and the virialisation parameter $K_{\rm vir}=(dV/dR)/(dV/dR)_{\rm vir}$ 
 which is a measure of the virialisation of the gas based on its velocity gradient. MyRadex is a revised version of RADEX
\citep{VanderTak07}, that fits the CO SLED with a large velocity gradient (LVG) model and adds as an additional constraint the CI flux value \citep{Papadopoulos+18}. It solves the statistical equilibrium problem with the LSODE\footnote{\href{https://ntrs.nasa.gov/citations/19940030753}{Livermore solver for ordinary differential equations}} solver in ODEPACK\footnote{\url{https://people.sc.fsu.edu/~jburkardt/f77_src/odepack/odepack.html}} and avoids unconverged solutions under some parameters compared to the original RADEX. We used the CO and CI molecular data from the Leiden atomic and molecular database \citep[LAMDA\footnote{\url{https://home.strw.leidenuniv.nl/~moldata/}};][]{Schoier05} and assumed background temperature as the cosmic microwave background (CMB) temperature at the cluster redshift, $T_{\rm back (z)}=T_{\rm CMB (z)}=3.69\rm \,K$.

We made a grid of numbers in the [$n_{H_2}$, $T_{\rm kin}$, $K_{\rm vir}$] parameter space, composed of $0.7<\rm{\log{\frac{T_{kin}}{[K]}}}<3$ with an increment of 0.11 dex, $1<\rm{\log{\frac{n_{H_2}}{[cm^{-3}]}}}<7$ with an increment of 0.1 dex, and $1<\rm{K_{vir}}<100$ with an increment of 0.1 dex. Then, we fitted a model for each point in our parameter space using MyRadex, and calculated the likelihood of each model based on the observations of CO and CI, following the method of \cite{Papadopoulos14}. The probability for each observation follows a Gaussian distribution, using $P={\rm exp}(-\chi^2/2)$, while $\chi^{2}$=$\Sigma_i
(1/\sigma_i)^{2}[R_{{\rm  obs}(i)}  -  R_{{\rm  model}(i)}]^2$,  where $R_{\rm  obs}$  is  the  ratio  of  the  measured  line  luminosities, $\sigma_{\rm i}$  the error  of the measured  line ratio,  and $R_{\rm model}$ the ratio of the  line brightness temperatures calculated by the  LVG  model. We used the \citwo\ observation as the additional constraint in our model. In particular, we used the ratio \citwo/\coone\ instead of the atomic Carbon ratio, \citwo/\cione, due to lack of the \cione\ line observation.

Using the definition of the line luminosity from Equation \ref{Lprime} and optically thin assumption, we can derive the \citwo\ luminosity following \citet{Papadopoulos+04} and \citet{Lei23} as
\begin{equation}
    L^{\prime}_{\rm model,CI(2-1)}=\frac{hc^3A_{21}X_{\rm CI}}{8\pi C\,\mu m_{\rm H_2}k\nu_0^2}Q_{\rm 21}M_{\rm H_2}
    \label{MH2CI}
\end{equation}

where $m_{\rm H_2}$ is the mass of a single H$_2$ molecule; $\mu = 1.36$ corrects for the mass in He; $C$ is the conversion between pc$^2$ and cm$^2$; $A_{21} = 2.68 \times 10^{-7} s^{-1}$ is \citwo\ spontaneous emission coefficient (Einstein coefficient); $X_{\rm C} = \rm [C]/[H_2]=3 \times 10^{-5}$ is the Carbon abundance \citep{Papadopoulos22}; $c$ is light speed; $k$ is the Boltzmann constant; $h$ is the Planck constant; $\nu_0$ is \citwo\ rest frequency; and $Q_{21}=N_{2}/N_{\rm CI}<1$ is the CI excitation factor depends on ($n_{\rm H_2}$, $T_{\rm kin}$) derived using equations from \cite{Papadopoulos+04}.
Using the LVG assumption, \cite{Papadopoulos+12b} derived CO conversion factor $\alpha_{\rm CO}=M_{\rm H_2}/L^{\prime}_{\rm CO(1-0)}$ as
\begin{equation}
    L^{\prime}_{\rm model,CO(1-0)}=\frac{\sqrt{\alpha}}{3.25}\frac{T_{\rm b,1-0}}{\sqrt{n_{\rm H_2}}}K_{\rm vir}M_{\rm H_2}
\end{equation}
where $\alpha = 1.5$ for typical cloud density profile, $T_{\rm b,1-0}$ is \coone\ brightness temperature, derived using LVG code depend on ($n_{\rm H_2}$, $T_{\rm kin}$, $K_{\rm vir}$).
Thus, we can derive $R_{\rm model,CI(2-1)}=L^{\prime}_{\rm model,CI(2-1)}/L^{\prime}_{\rm model,CO(1-0)}$ and add this as an additional constraint to the $\chi^2$ calculations.

Figures ~\ref{fig:BCG_model} and ~\ref{fig:CGM_model} show the best-fitted models for the BCG and the tail, respectively. The top panels of both figures show the distribution of the models in the \Tk, $n_{H_2}$ space where the colours show the \citwo/\coone\ ratio of the model that has the highest likelihood between models with the same \Tk, $n_{H_2}$, but different $\rm K_{vir}$. Black contours show the levels of \citwo/\coone\ and white contours show the likelihood levels. The white star represents the best-fitted solution based on observations. 
The bottom panels of both Figures ~\ref{fig:BCG_model} and ~\ref{fig:CGM_model} show the CO SLED of the models compared to observations. The black curve represents the best-fitted model with the highest likelihood. The blue points show the CO line observations. The green curves show the scatter of the best models after applying the Markov chain Monte Carlo (MCMC) sampling to the solutions. We set wide parameter range in the MCMC sampling by assuming flat priors for $5<\rm{\frac{T_{kin}}{[K]}}<10^3$, $1<\rm{\log{\frac{n_{H_2}}{[cm^{-3}]}}}<7$, $1<\rm K_{vir}<100$ in logarithmic scale with prior probability $P(p)=1$, and set $P(p)=0$ for solutions outside the prior criteria \citep{Papadopoulos14}. The best models predict the gas properties of \Tk=$22.6_{-4.3}^{+8.3} $K, $\rm{\log{\frac{n_{H_2}}{[cm^{-3}]}}}=4.5_{-0.8}^{+1.3}$, and $\rm K_{vir}=4.6_{-3.0}^{+15.0}$ for BCG and \Tk=$15.3_{-5.3}^{+15.6} $K, $\rm{\log{\frac{n_{H_2}}{[cm^{-3}]}}}=3.5_{-1.6}^{+2.2}$, and $\rm K_{vir}=19.0_{-14.2}^{+83.5}$ for tail. Based on the current data and models, we argue that the tail gas might be cooler, less dense, and less bound compared to the BCG gas, on average. Considering that the CR energy density in a star-forming galaxy with strong AGN activity is expected to be higher than the CR energy density in Milky-Way, kinetic temperatures below $\sim 20$ K for the tail gas are difficult to explain theoretically (See Figure 2 of \citealt{papadopoulos11}). 
It is noteworthy that the top panel of Figure ~\ref{fig:BCG_model} shows two possible solutions for the BCG (white contours), one colder and denser and one hotter and more diffuse with lower probability. However, a temperature of more than 1000 K is not physical for the BCG gas. Thus, we do not consider the second solution as a valid model. This degeneracy in the [$n_{H_2}$-$T_{\rm kin}$] space is inherent to this kind of models and can only be broken with additional independent observations, as the ratio of the two atomic carbon lines, \cione/\citwo\, and a better sampling of the CO ladder. Figure \ref{fig: corner} in Appendix \ref{appendix-B} presents the joint distribution of the model parameters for all MCMC samplings. These plots show that the physical conditions of the BCG are better constrained compared to the CGM tail. Moreover, the model predictions for the physical conditions of these two galaxy components are sufficiently separate by $2\sigma$ in the $n_{H_2}$-$T_{\rm kin}$ plane. However, $\rm K_{vir}$ is not well constrained for any of the galaxy components.



\section{Discussion}\label{discussion}

\subsection{CGM tail formation and excitation}

Our models in Section \ref{modeling} show that the BCG of MACS1931 has the kinetic temperature of \Tk=$22.6_{-4.3}^{+8.3} $K, the molecular hydrogen number density of  $\rm{\log{\frac{n_{H_2}}{[cm^{-3}]}}}=4.5_{-0.8}^{+1.3}$, and the virialistion parameter of $\rm K_{vir}=4.6_{-3.0}^{+15.0}$. The modelled kinetic temperature of the detected gas in BCG is less than the dust temperature of 33 K, estimated by \citet{Santos+16} using the Herschel photometry, but because of the large error bars we may conclude that the two temperatures are similar and the BCG gas is thermalised.

In the MACS1931 CGM tail, we detect gas with a low kinetic temperature, \Tk=$15.3_{-5.3}^{+15.6} $K, a molecular hydrogen number density of $\rm{\log{\frac{n_{H_2}}{[cm^{-3}]}}}=3.5_{-1.6}^{+2.2}$, and a virialistion parameter of $\rm K_{vir}=19.0_{-14.2}^{+83.5}$. In this case, one could conclude that \Tk~ > \Td~ $\sim 10$\ K \citep{Fogarty19} and the gas might not be in LTE conditions \citep{Papadopoulos22} with the \citwo\ gas in a subthermal state. However, in addition to the large uncertainties of the \Tk\ measurements, the value of \Td\ estimated by \citet{Fogarty19} is also affected by the uncertainties related to the dust model and the subsequent value assumed for the dust emissivity index.

For the CGM tail of MACS1931, the high \cothree/\coone\ brightness temperature
ratio (${\rm R_{31}=\rm \frac{I_{CO(3-2)}}{I_{CO(1-0)}}\times
\frac{1}{9}=0.67}$, Table~\ref{tab:lines})
is consistent with both cold and dense as well as warmer (and less dense)
molecular gas in the known number density-kinetic temperature ($\rm n_e$-\Tk ) degeneracy. This is inherent in the radiative
transfer modelling of optically thick lines like those of \CO. However, this high brightness temperature ratio is
typically associated with the warm  ISM found in starbursts. Indeed the
observed \cothree/\coone\ ratio (R$_{31}$) in the CGM tail is identical to the mean value found for the star-forming ISM of Luminous Infrared Galaxies (LIRGs)
\citep[see Figure 5 in ][]{Papadopoulos+12}.
In their study, additional \coone\ and
$^{13}$CO line observations limit the $\rm n_e$-\Tk\ degeneracy in the LIRGs studied,  attributing such high \cothree/\coone\ ratios to the warm end of the $\rm n_e$-\Tk\ parameter space.
Figure~\ref{fig:CGM_model} shows that the predicted CO SLED of the tail models can peak at different $\rm J_{upper}$ values. But because of the large error bars on the existing data and the lack of higher-$\rm J_{upper}$ CO transitions the CGM gas status cannot be pinned down consequently.

The BCG and CGM of MACS1931 contain warm molecular hydrogen, observed in the preliminary data analysis of a cycle 2 MIRI/MRS program with JWST (program ID: 3629, PI: A.Man). Mapping and modelling the JWST warm hydrogen data and the ALMA cold hydrogen data in this system will shed new light on the multiphase nature of the CGM gas and provide clues on the role of the CRs in the excitation of CGM and it is the subject of a following paper.
Confirming a warm \HII gas phase concomitant with such a cold dust using additional molecular, atomic line, and dust continuum observations, would verify the thermal decoupling of the two phases of the gas, and the role of CRs in warming the gas (but not the dust) rather than the usual star formation-originating far ultraviolet(FUV) radiation fields. In particular, the R$_{31}$ ratio seems to be sensitive to an enhanced CR energy density \citep{Bisbas+23}. Here we must mention that in principle turbulence can also warm the gas but not the dust in terms of the global energy budget necessary. However, the detailed picture emerging from simulations as well as observations is that only a small fraction ($\sim $1-2\%) of a supersonically turbulent gas reservoir is heated to extraordinarily high temperatures ($\sim $1500-2000\,K) while the bulk remains very cold \citep[$\sim $10-15\,K; ][]{Wu17, Bisbas+21}.

\citet{Fogarty19} shows that the tail direction is not aligned with the AGN jet axis. They argue that this tail-like structure is likely an infalling feature which has been ejected out of the galaxy by AGN jets in an outburst long ago. This structure is currently falling back into the galaxy and fueling the active star formation of MACS1931 BCG due to a more recent AGN outburst which uplifted the gas and now the gas is condensed and raining back to the galaxy. \citet{Ciocan21} studies the ionised and molecular gas phases of MACS1931 using VLT-MUSE optical integral field spectroscopy and archival ALMA observations. They show that both CO and $\rm H\alpha$ emission are co-spatial and share similar kinematics. This study highlights that the gas phase metallicity measured form optical lines in the CGM tail is more than that in the BCG. Their findings are in agreement with the scenario suggested by \citet{Fogarty19}. \citet{Ehlert11} finds two bright ridges about 25-30 kpc to the north and south direction of the BCG using \textit{Chandra} X-ray observations. This study probes gas with a temperature of $10^7-10^8 \rm K$ and a density of $10^{-3}-10^{-1} \rm cm^{-3}$. The northern ridge, which is co-spatial with the CGM component as seen in the ALMA observations, is the coldest ($k_BT=4.78 \pm 0.64 \, keV$), densest ($\rho \sim 0.09 \rm \, cm^{-3}$), and the most metal-rich ($Z = 0.53 \pm 0.11 \, Z_{\odot}$) region in the cluster. They argue that these ridges originally formed as a result of a previous cluster merger and an AGN outburst in the BCG. Furthermore, \textit{Subaru} optical observations presented in \citet{Ehlert11} show continuum emission from a young stellar population in both north and south of the BCG and $\rm H\alpha$ emission from star-forming regions only in the north of BCG, almost coinciding with the north X-ray ridge. They suggest that the metal-rich cool core of the cluster has been stripped and moved to the north of the BCG as a result of a previous cluster merger. The tail-like structure detected in ALMA observations lies approximately in the northern X-ray ridge and the star-forming region detected in optical. 
Our results show that the molecular gas in the tail is colder than the molecular gas in the BCG, consistent with the condensation scenario. However, we need more information to understand whether the dynamical state of cluster post-merger affects the star formation of the BCG.


\subsection{Molecular hydrogen mass inferred from \citwo }\label{gasmass}

An estimate of the molecular hydrogen mass in MACS1931 can be obtained by rearranging Equation \ref{MH2CI} as

\begin{equation}
M_{H_2} [M_{\odot}] = 1375.8 \frac{D_L^2}{1+z} (\frac{X_{CI}}{10^{-5}})^{-1}  (\frac{A_{21}}{10^{-7}})^{-1} Q_{21}^{-1} \frac{S_{\citwo}\Delta V}{\rm Jy\,km\,s^{-1}}
\label{eq:H2CI}
\end{equation}

Where $D_L$ is the luminosity distance to the object equal to 1557 Mpc; $z=0.3525$ is the redshift of the object; $S_{\citwo}\Delta V$ is the flux of \citwo~reported in Table \ref{tab:fluxes}. We calculate the probability distribution of this quantity from \citet{Papadopoulos+04} for the BCG of MACS1931 with \Tk=$22.6_{-4.3}^{+8.3} $K and $\rm{\log{\frac{n_{H_2}}{[cm^{-3}]}}}=4.5_{-0.8}^{+1.3}$ to be $Q_{21}=0.13 \pm 0.05$. Equation \ref{eq:H2CI} gives $\rm M_{H_2}\rm{[BCG] = (10.0 \pm 3.2) \times 10^{10} M_{\odot}}$ as the molecular hydrogen mass in the BCG of MACS1931, traced by \citwo\ if we only consider the measurement errors. We perform the same analysis for the CGM tail of MACS1931, using $Q_{21}=0.03 \pm 0.02 $ for gas with \Tk=$15.3_{-5.3}^{+15.6} $K and $\rm{\log{\frac{n_{H_2}}{[cm^{-3}]}}}=3.5_{-1.6}^{+2.2}$. Using the tail flux and its uncertainty from Table \ref{tab:fluxes}, we find $\rm M_{H_2}\rm{[CGM]=(13.4 \pm 9.6) \times 10^{10} M_{\odot}}$ as the molecular hydrogen mass in the tail of MACS1931. Consequently, the total molecular hydrogen mass of MACS1931 in BCG and the tail is $\rm M_{H_2}\rm{[tot]=(23.4 \pm 12.7) \times 10^{10} M_{\odot}}$ based on \citwo\ observations. In this calculation, we only consider the measurement uncertainties on the line fluxes. However, a larger uncertainty lies within the $Q_{21}$ factor and the final result significantly depends on this factor, specifically for the tail. For instance, using the upper limit of the $Q_{21}$ factor for tail, the hydrogen mass is reduced by $\sim 30 \%$.

The lower limit of this estimation is higher than the hydrogen mass of MACS1931 as $M_{H_2}\rm{[tot]} = 1.9 \pm 0.3 \times 10^{10} M_{\odot}$ traced by \coone\ ALMA observations \citep{Fogarty19}. This comparison shows that part of the total molecular gas in BCG and CGM might be CO-poor or CO-dark, but it can be detected using \citwo. Consequently, we find $\sim 10$ times higher gas-to-dust ratio for MACS1931 which might alleviate the tension of the low gas-to-dust ratio as reported in \citet{Fogarty19} and nominal Galactic values.


\section{Conclusion}

In this work, we present new single-dish observations of the brightest cluster galaxy and its circumgalactic medium belonging to the cool-core galaxy cluster, MACS 1931-26.
The data obtained by the APEX/SEPIA660 receiver reveals \coseven\ emission from the BCG and \citwo\ emission lines originated from BCG and the CGM tail. The NRO 45-m/FOREST observation shows a \coone\ emission line from the BCG. The main conclusions of this work are:

\begin{itemize}

    \item 
    The \coone~flux derived from the NRO 45-m single-dish observation of the MACS1931 BCG is consistent with the \coone~flux of this object calculated in \citet{Fogarty19} using ALMA interferometry data. Thus, the ALMA observations are not missing the flux from large-scale structures of this object and the low gas-to-dust ratio of this system is not an interferometric artefact.

    \item 
    The LVG models show that the MACS1931 BCG molecular gas is excited with \Tk=$22.6_{-4.3}^{+8.3} $K, $\rm{\log{\frac{n_{H_2}}{[cm^{-3}]}}}=4.5_{-0.8}^{+1.3}$, and $\rm K_{vir}=4.6_{-3.0}^{+15.0}$. The BCG gas might be warmer, denser, and more bound compared to the CGM tail molecular gas with \Tk=$15.3_{-5.3}^{+15.6} $K, $\rm{\log{\frac{n_{H_2}}{[cm^{-3}]}}}=3.5_{-1.6}^{+2.2}$, and $\rm K_{vir}=19.0_{-14.2}^{+83.5}$. 

    \item 
    The lower limit of the molecular hydrogen mass estimated from the \citwo\ emission ($>10.7 \times 10^{10} M_{\odot}$) is larger than the hydrogen mass calculated using \coone\ in \citet{Fogarty19}, suggesting that part of the molecular gas in this system might be CO-poor and the gas-to-dust ratio might be higher than the value calculated previously using \coone\ observations only. Using a combination of \CO\ and CI observations lets us trace the different phases of molecular gas including the CO-poor regions.

\end{itemize}

Detecting the CGM in emission is complicated because the emission is rather faint and very extended, making it difficult with (sub)mm interferometers like ALMA, which are currently the (sub)mm observatories with the most collecting area. Significant progress in mapping the CGM in emission will likely require a large single-dish telescope such as the Atacama Large Submm Telescope \citep[AtLAST;][]{Lee24}.

Observations of \cione\ and higher CO transitions are needed to constrain the properties of the excited tail gas and investigate the existence of CRs in this environment. For instance, the \cofive\ line with an excitation temperature of $\sim$83 K would add useful constraints on high-J/low-J CO line ratios and the type of  the gas heating mechanisms in this system.


\begin{acknowledgements}
We express our sincere gratitude to Padelis Papadopoulos for his invaluable contributions to our discussions and for providing insightful feedback on this work which significantly enhanced the quality of this project.
LG and AWSM acknowledge the support of the Natural Sciences and Engineering Research Council of Canada (NSERC) through grant reference number RGPIN-2021-03046.
PA warmly thanks the Department of Physics, Section of Astrophysics, Astronomy and Mechanics, Aristotle University of Thessaloniki for its hospitality during her sabbatical time and where this work has been developed. PA and YM acknowledge the support of Invitational Fellowships for Research in Japan.
AWSM is grateful for the Hellenic captain, whose hospitality and boundless mind prepared the sails for this maiden voyage to explore uncharted waters and shall continue until all oceans are explored.
This article makes use of data from ALMA program ADS/JAO.ALMA\#2016.1.00784.S and ADS/JAO.ALMA\#2017.1.01205.S. ALMA is a partnership of ESO (representing its member states), NSF (USA) and NINS (Japan), together with NRC (Canada) and NSC and ASIAA (Taiwan) and KASI (Republic of Korea), in cooperation with the Republic of Chile. The Joint ALMA Observatory is operated by ESO, AUI/NRAO and NAOJ.
This publication is based on data acquired with the Atacama Pathfinder Experiment (APEX) under programme ID E-0106.A-0997A-2020. APEX is a collaboration between the Max-Planck-Institut fur Radioastronomie, the European Southern Observatory, and the Onsala Space Observatory
The Nobeyama 45-m radio telescope is operated by Nobeyama Radio Observatory, a branch of National Astronomical Observatory of Japan. JZ and ZZY acknowledge the support of the National Natural Science Foundation of China (NSFC) under grants No. 12041305, 12173016, the Program for Innovative Talents, Entrepreneur in Jiangsu, and the science research grants from the China Manned Space Project with NOs.CMS-CSST-2021-A08 and CMS-CSST-2021-A07.

\end{acknowledgements}

%
%
\bibliographystyle{aa}
\bibliography{bib.bib}



\begin{appendix}
\section{Joint distribution of the LVG model parameters}
\label{appendix-B}

\begin{figure*}
\hspace{1.2cm}
\includegraphics[width=15cm]{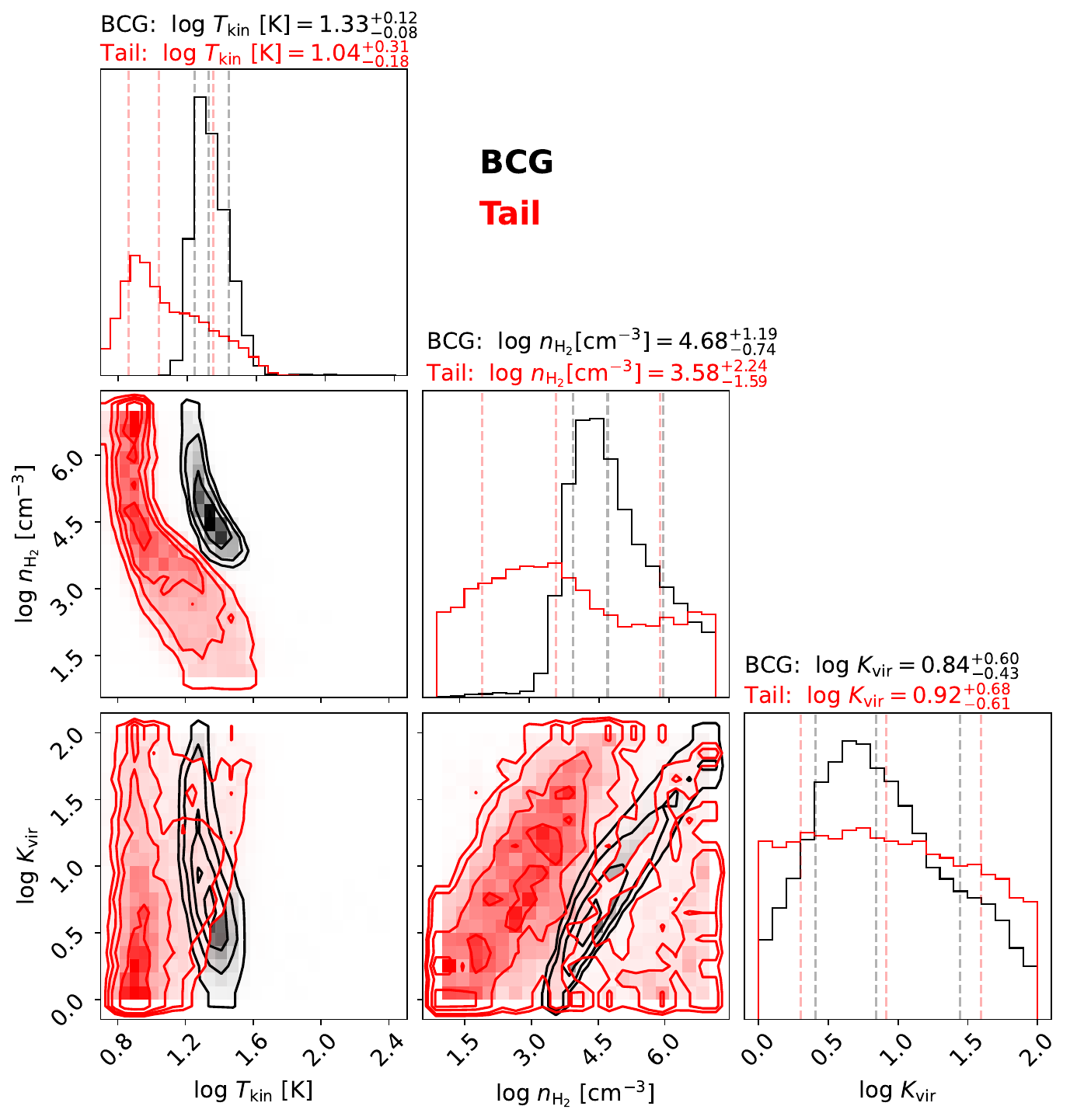}
\caption{Corner plot of the model parameters of the MCMC result. Contours and histograms of BCG and CGM are shown in black and red, respectively. The off-diagonal panels show the joint distriburtion of each two parameters for BCG and CGM with contours showing $[0.5, 1,1.5,2] \sigma$ levels. The diagonal panels show the histograms of all parameters including $\rm \log{\frac{T_{kin}}{[K]}}$ on the top left panel, $\rm{\log{\frac{n_{H_2}}{[cm^{-3}]}}}$ on the central panel, and $\rm \log{K_{vir}}$ on the lower right panel. The 16\%, 50\%, and 84\% percentiles are shown with dashed black and red lines for BCG and CGM, respectively. These percentile values are reported on top of each histogram panel.}
 \label{fig: corner}
\end{figure*}

The joint distribution of the model parameters resulted from MCMC sampling on both BCG and the CGM tail are shown in Figure \ref{fig: corner}. These plots show that the BCG and the CGM tail are substantially distinguishable in the temperature-density plane. However, $\rm K_{vir}$ is not a distinctive parameter for the BCG and the CGM tail. (Section \ref{modeling}).

\end{appendix}

\end{document}